%Paper: hep-th/9308038
%From: "V. Petkova" <PTVP@IBM.RZ.TU-CLAUSTHAL.DE>
%Date: Sun, 08 Aug 93 23:26:25 MET
%Date (revised): Tue, 21 Sep 93 18:04:34 MET

\magnification 1100
\font\male=cmr8

\def\za{\alpha} \def\zb{\beta}  
\def\ze{\varepsilon}   \def\zk{\kappa}
\def\zl{\lambda}   
 \def\zr{\rho}

 \def\zD{\Delta}

       \def\cL{{\cal L}}
     \def\cO{{\cal O}}  
   \def\cR{{\cal R}}    
     \def\cW{{\cal W}}

\def\Ri{{\cal R}^{(i)}}
\def\odots{{\scriptstyle{\circ\atop\circ}}}
\def\p{\partial}

  \def\C{C}
\def\kk{{1\over\nu}}
\def\Tf{T^{(\rm f{}f)}} \def\Wf{W^{(\rm f{}f)}}
\def\Lf{L^{(\rm f{}f)}}
\def\aw{a_w}

\def\hh{\hat h} \def\hf{\hat f} \def\he{\hat e}
\def\dZ{Z\!\!\!Z}
\def\s3{sl(3)} \def\s2{sl(2)}
\def\V{V_{\lambda}}

\def\W{{\cal W}}
 
\def\GP{G^+_{0}} \def\GM{G^-_{-1}}

\def\tp{t_+} \def\tm{t_-} \def\t0{t_0} \def\tpm{t_\pm}

% caligraphic:

% doubled letters:

 \def\dN{I\!\!N}      
     \def\dZ{Z\!\!\!Z}

\def\la{\langle} \def\ra{\rangle}

\def\pb{{\rm (PB)}}   \def\ol{\overline\lambda}

\nopagenumbers

\centerline{ }

\hfill
 SISSA -- 106/93/EP

\vskip 2cm
\centerline{ ON  $\widehat{sl}(3)$ REDUCTION, QUANTUM
 GAUGE TRANSFORMATIONS,}
\centerline{ AND  $\W -$ ALGEBRAS
 SINGULAR VECTORS }
\vskip 2cm

\centerline{\bf P. Furlan}
\medskip
\centerline{ Dipartimento di Fisica Teorica
             dell'Universit\`a di Trieste, Trieste,}
\centerline{ and Istituto Nazionale di Fisica Nucleare,
             Sezione di Trieste, Italy}

\bigskip
\centerline{\bf A.Ch. Ganchev\footnote{$^\sharp$}
                              {{\male %Postdoctoral fellow of
                              % Istituto Nazionale di Fisica Nucleare,
                              % Sezione di Trieste, Italy;
                               Permanent address: INRNE,
                               Sofia, Bulgaria.}}  }
\medskip
\centerline{ Istituto Nazionale di Fisica Nucleare,
                              Sezione di Trieste,}
\centerline{ Scuola Internazionale Superiore di Studi Avanzati,
             Trieste, Italy }

\bigskip
\centerline{\bf V.B. Petkova \footnote{$^*$}
{{\male Address after ~August 1, 1993 : Arnold Sommerfeld Institute
f. Math. Physics, Tech. Univ. Clausthal, 38678
Clausthal-Zellerfeld, Germany.}} }

\medskip
\centerline{ Istituto Nazionale di Fisica Nucleare, Sezione di Trieste,
Italy, }
\smallskip
\centerline{  Institute for Nuclear Research and Nuclear Energy,
Sofia 1784, Bulgaria \footnote{$^{**}$}{{\male Permanent address.}} }
\vskip 2cm

\centerline{\bf Abstract}
\vskip 1cm

The problem of describing the singular vectors
 of $\cW_3$ and  $\cW_3^{(2)}$ Verma modules is addressed,
 viewing these algebras as BRST quantized  Drinfeld-Sokolov (DS)
reductions of  $A^{(1)}_2\,$.
Singular vectors of an $A^{(1)}_2\,$ Verma module are mapped
into $\W$ algebra singular vectors and
are shown to differ from the latter
by terms trivial in the BRST cohomology.
These maps are realized by quantum versions of the highest weight
DS gauge transformations.
\vskip 1cm

\vfill July 1993

\vfill\eject
\pageno=1
\footline{\hfil\tenrm\folio\hfil}

\bigskip\noindent{\bf 1.\ }
The quantum DS reduction of $\widehat{sl}(2)$ Verma modules
to Virasoro Verma modules was considered by two of the authors in
[1]. The main tool in that work was a kind of quantum gauge
 transformation which translates, in an explicit fashion,
the singular vectors of $\widehat{sl}(2)$ into Virasoro singular vectors.
This letter is a generalization of the method of [1]
to the case of $\widehat{sl}(3)$.

There are two different reductions of $\widehat{sl}(3)$ --
to the $\W_3$ algebra [2] of Zamolodchikov (Z) or the $\W^{(2)}_3$
algebra [3] of Polyakov-Bershadsky (PB), best described in the
general scheme of de Boer-Tjin [4].
After setting the notation and
emphasising the connection between the classical
highest weight DS gauge [5], [6] and the work [4]
we proceed to define the relevant quantum gauge transformations.
With their help we illustrate on a few examples how the reduction
of the  singular vectors of Verma modules is accomplished.
In particular we recover the
subclass of $\W_3$ singular vectors obtained in [7] and
its counterpart for  Verma modules of $\W^{(2)}_3$.
The details are left for a
 more systematic work  in preparation.
For general information we refer to the reviews [8], [9].

\bigskip\noindent{\bf 2.\ }
Here we set some notation and
recall a few facts about the affine Lie algebra
$A_2^{(1)}$ (see [10],[11]).

The positive roots of $A_2$ are the simple ones
$\alpha_1$, $\alpha_2$ and the highest root
$\alpha_3 \equiv \alpha_1+\alpha_2$. The Cartan -- Killing form
on  $A_2$  is $\la X,Y \ra =tr (XY) $. Denote
 by $\left( \C^{ij} \right)_{i,j=1,2}=
\la\za_i,\za_j\ra$ the Cartan matrix and
by  $\left( \C_{ij} \right) $ -- its inverse .
The Cartan-Weyl basis consists of
$e^a$, $f^a$, $a=1,2,3$, and  $h^i= [e^i,f^i]$, $i=1,2$, generating
the subalgebras  $ {\tt n_+ }\,,$ $ {\tt n_- }\,$ and ${\tt h}$  of $A_2
 ={\tt n_+}\oplus {\tt h} \oplus {\tt n_-} $ respectively.

The algebra
 $A^{(1)}_2$ with  the derivation ${\tt d}$ added
  admits the decomposition ${\tt N_+} \oplus {\tt  H} \oplus {\tt N_-}$ where
 ${\tt H}$ is the Cartan subalgebra consisting of ${\tt d}$,
 {\tt h} and the central element, which will be assumed to be a
fixed complex number $k$,  different from $-3$.
The subalgebras ${\tt N_+}$ and ${\tt N_-}$ are generated by
$\{e^1_0, e^2_0, f^3_1 \}$ and $\{f^1_0, f^2_0, e^3_{-1}\}$ respectively.
 A $A_2^{(1)}$ Verma module is built by the action of ${\tt N_-}$ on a highest
 weight state $V_{\zl}$, annihilated by the elements of ${\tt N_+}$.
 The projection of the weight $\zl$ on ${\tt h}^*\,$
 %the (dual) Cartan subalgebra of $A_2$
is denoted by $\ol\,$; $\, \la \ol , \za_1 \ra + \la \ol , \za_2 \ra +
 \la \zl , \za_0 \ra = k \,.$
Denote $M_i=\la \zl + \rho, \za_i \ra$, $i=1,2,3\,,$
and $\nu^{-1} = k + 3\,$;
%$\rho = 3 \zL_0 +\za_1 +\za_2$.
$\la \rho, \za_j \ra=1\,,$ $j=0,1,2$.

If some of the conditions
$$
  M_i = \cases{ \pm \Big( m_i - {n_i\over\nu}\Big)\, & $m_i, n_i\in\dN$\cr
                m_i                                   & $m_i \in \dN$ \cr}
\eqno(1)$$
$i=1,2,3$, hold, then the Verma module of highest weight $\zl$ is reducible.
The highest weights of the embedded modules are obtained by the shifted
action of the affine Weyl group on $\zl$
(to be denoted by $w_{\za} \cdot \zl$) . In particular
the simplest series of singular vectors correspond to the simple
roots $\za_1\,, \za_2\,, \za_0$ (i.e., $\la\zl + \zr , \za\ra = m_1\,
, m_2\,,$ or, $ 1/\nu - M_3\,, $
respectively, are positive integers). Explicitly
the corresponding singular vectors are given by
$$
  V_{w_{\za_i}\cdot \zl} =
          (f^i_0)^{m_i}\,  \, \V \,, \quad i=1,2\,,
  \qquad\quad V_{w_{\za_0}\cdot \zl} =
          (e^3_{-1})^{{1\over \nu} -M_3}\,  \, \V \,.
\eqno(2)$$

In general decomposing the relevant element
of the affine Weyl group into simple reflections $w_0$, $w_1$, $w_2\,,$
 ($ w_{j} \equiv  w_{\za_j}\,, j=0,1,2\,$ ),
  one can write down expressions for the singular vectors [11].
As an illustration consider the weights for which
 $M_1=m-\kk$, $m\in\dN$. Then there is a singular vector
corresponding to the root $\za= \za_0+\za_3+\za_1$
 $$
V_{w_{\za}\cdot \zl}=  V_{w_1 w_0 w_2 w_0 w_1\cdot \zl} =
     (f^1_0)^{m+\kk} \,  (e^3_{-1})^{m-\kk+M_2} \, (f^2_0)^m \,
     (e^3_{-1})^{\kk-M_2} \,  (f^1_0)^{m-\kk} \, \V \,.
\eqno(3)$$
The above formula is a monomial of the generators raised, in general,
to complex powers acting on $\V$. Nevertheless these monomials
can be rewritten as ordinary (integer power) polynomials of the
elements of the  subalgebra ${\tt N_-}$.
Note that formally the first (counted from the right),
the first two, the first three, etc.,
factors of these monomials also give singular vectors,
but only formally, since the corresponding operators are not
elements of the universal enveloping algebra.

For the elements of the chiral algebras we will use both the
notation in terms of modes $A_n$ and currents
$ A(z) = \sum_{n\in\dZ} A_n \, z^{-n-\zD_A} $.
We assume that all Kac-Moody currents $X(z)$ have $\Delta_X=1$.
The (anti)commutation relations of the modes are equivalent to
the singular part of the operator product expansion (OPE).
The normal product of two fields $(A\,B)(z)$ will be defined as the
zero order term in the expansion of $A$ and $B$.
We have used the computer program of [12] for the more tedious OPE
computations.

\bigskip\noindent{\bf 3.\ }
We briefly
recall the classical situation [5], [6]. One considers the matrix
operator
$ \zk \, \p_z + A(z) $ with
$
  A(z) =  f^a(z)\,\tp^a + e^a(z)\,\tm^a +
          C_{ij} \, h^i(z) \, \t0^j
\,,$
where $\tpm^a$, $a=1,2,3$ and $\t0^i$, $i=1,2$ is the Cartan-Weyl
basis of $sl(3)$
 and summation over repeated indices is assumed.
The functions $e^a(z)$, $f^a(z)$, and $h^i(z)$
 together with the constant $\zk $
  are coordinates
 on the dual $\widehat{sl}(3)^*$ of $\widehat{sl}(3)$ and they
close a classical (Poisson bracket)
KM algebra.
The first step in the Hamiltonian reduction of the classical KM algebra
is to impose a set of first class
 constraints on the $e$'s, corresponding to a
subalgebra $ {\tt n}^0\, $ of the  nilpotent algebra ${\tt n_+}$.
The second step is to factor out the gauge group generated by
these constraints. The result is a new phase space with coordinates
described by gauge invariant functions on the constrained space and
Poisson bracket inherited from that on  $\widehat{sl}(3)^*$ .

The action of the group is given by the coadjoint
 representation
 $ A_g(z) = g^{-1}(z) \, A(z) \, g(z) + \zk g^{-1}(z)\, \p g(z)\,, $
$ \,g(z) = 1 + \sum
\, b^a(z) \, \tp^a\,,$
 where the sum is  over $ \tp^a \in {\tt n}^0\,.$
The factorisation is done by fixing the gauge, i.e., performing
 a  gauge fixing transformation with
 parameters
$b^a(z)$ being proper functions of the   unconstrained
currents and their derivatives.

Let us recall the two ways of reducing $\widehat{sl}(3)^*$, corresponding to
the two inequivalent  embeddings of $sl(2)$ into $sl(3)$.
In the case of the principal embedding the constraints are
$$
  e^1(z)=e^2(z)=1, \qquad e^3(z) = 0 \,.
\eqno(4)$$
There are three gauge fixing conditions and hence two surviving
gauge invariant functions. This corresponds to the splitting of
the adjoint representation of $sl(3)$ into a spin $1$ and a spin $2$
representation of $sl(2)$.
Choosing the highest weight gauge  one has
$h^i \rightarrow 0\,,$  $f^1 - f^2 \rightarrow 0\,,$
$f^1 +f^2 \rightarrow u_1\,,$ $\ f^3 \rightarrow u_2\,,$
with the gauge invariant functions $u_i = u_i (f^a, h^j)\,$ being
$$\eqalign{
  u_1 &= f^1+f^2 + u^{\rm (f{}f) }_1
  = f^1 +f^2 + {C_{ij}\over 2} h^i h^j + \zk\, \p h^3  \,,
\cr\cr
  u_2 &= f^3 + {\zk\over 2}\p (f^1-f^2) +
  C_{2j}h^j\, f^1  -  C_{1j}h^j\,f^2 + u^{\rm (f{}f) }_2 \,,
\cr\cr
   u^{\rm (f{}f) }_2 &=
  C_{1i}C_{2j} (C_{1l}-C_{2l})h^l\, h^i\, h^j
             \ + {\zk\over 2} \,
              \left( C_{1i}\,  h^i\, \p h^1 -
              C_{2i}\,  h^i\,\p h^2 \right)
  + {\zk^2\over 6} \,\p^2 (h^1-h^2) \,,
\cr} \eqno(5)$$
thus providing the generating functions of the
 classical analogue of the Zamolodchikov
$\cW_3$ algebra (the (Z) case for short).
The gauge fixing transformation is
$$
  g(z) = 1 + \sum_{a=1}^3 \,b_a(z)\, t_+^a \, \,, \qquad {\rm with}\quad
  b_i(z)=C_{ij}\,h^j(z)\,, \quad i=1,2\,,
\eqno(6)$$
and $b_3(z)$  is a function of $f^i\,, h^i\,, \p h^i
 \,, i=1,2\,,$ which we will not specify.

%$b^3(z)= (f^1-f^2+(C_{1j}h^j)^2 - (C_{2j} h^j 2\, \p h^2 +
%C_{1j}h^j\,C_{2j}h^j +\zk \p (C_{1j}h^j-C_{1j}h^j) )/2$

The other embedding gives the
classical analogue of the $\cW_3^{(2)}$ algebra  (the (PB) case).
The constraints are
$$
  e^3(z)=1, \qquad e^2(z) = 0 \,.
\eqno(7)$$
Applying the highest weight gauge transformation
$$
  g^{\pb}(z) = 1 +  e^1(z) \, t_+^2 +
 {h^3(z)\over 2} \, t_+^3\,,
\eqno(8)$$
one gets $e^1  \rightarrow 0\,,$ $h^3  \rightarrow 0\,,$ and
$$\matrix{
    h^1-h^2 &\rightarrow  &u_0 =h^1-h^2 \,, \qquad
  & f^2 &\rightarrow & u_{1/2}^- = f^2 + e^1 h^2 +\zk \p e^1 \,,\qquad \cr\cr
    \quad f^1 &\rightarrow &u_{1/2}^+ =f^1 \,, \qquad\qquad
  & f^3 &\rightarrow & u_1^{\pb} = f^3 +e^1 f^1 + {1\over 2} C_{ij} h^i h^j
    + {\zk\over 2} \p h^3 \,. \cr}
\eqno(9)$$
The four gauge invariant polynomials $u_i$
(corresponding to the four $sl(2)$ representations -- of spin
$0$, ${1\over 2}$,  ${1\over 2}$, and $1$,
in the adjoint representation of $sl(3)$)
generate the classical (Poisson) algebra $\cW_3^{(2)}$.

\bigskip\noindent{\bf 4.\ }
In the BRST formalism one needs a pair of fermionic ghost fields
$b^a,c^a$ for each constraint. They have OPEs
$ b^a(z) \, c^b(w) = { \delta_{a,b} \over z-w} + \dots \,$
and $\Delta_{b^a} + \Delta_{c^a}=1$.
We choose $\Delta_{b^a}=0\,$ and $\Delta_{c^a}=1$.

First we will consider the reduction leading to the $\W_3$ algebra
of Zamolodchikov (Z) [2] associated to the principle
embedding.
The BRST charge is
$$
  Q =\sum_{a=1}^3 (e^{a}\,c^a)_{-1}
      - (b^3 (c^1 \,c^2))_{-1} - c^1_0 - c^2_0 \,.
\eqno(10)$$

Following the general scheme of [13], [4] let us introduce
 the ``hatted''  currents
$\hat X^a(z) = X^a(z) + f^{a \za }_{\zb}\,(b^{\zb}\ c_{\za})(z)\,,$ where
the summation indices $\za, \zb$ correspond to the constrained generators
$e^{\za}\,, \za= 1,2,3$. Explicitly
$$\eqalign{
  \hf^{1} & = f^{1} + (b^2\,c^3)\,, \qquad
  \hf^{2}   = f^{2} - (b^1\,c^3)\,, \qquad
  \hf^{3}   = f^{3}\,,              \cr
  \he^1   & = e^1+(b^3\,c^2)    \,, \ \qquad
  \he^2     = e^2-(b^3\,c^1)    \,, \qquad
  \he^3     = e^3               \,, \cr
  \hh^1   & = h^1 + 2(b^1c^1) - (b^2c^2) + (b^3c^3) \,, \qquad \qquad
  \hh^2     = h^2 + 2(b^2c^2) - (b^1c^1) + (b^3c^3) \,. \cr}
\eqno(11)$$
The OPEs among the
fields $\hf,\hh$ are the same as among the corresponding
unhatted ones with the only difference that $k$ is shifted to $k+3$.

The reduced currents $\,T(z)\,, W(z)\,,$ commuting with $Q$
 can be obtained as a ``quantization'' of
the gauge invariant differential polynomials (5).
%$$
%  T(z) = \nu \, :u_1(z): \,, \qquad
%  W(z) = {1\over\aw} \, :u_2(z): \,,
%\eqno(rgz)$$
Namely, substitute all generators in (5) by their hatted
counterparts (11), normal order the products (i.e., replace $h^j(z)
f^1(z)\,$ by $\, (\hh^j \hf^1)(z)\,, $ etc.),   and identify
$\zk = {1\over\nu}-1 = k+2\,.$ In modes this gives
$$\eqalign{
  {1\over  \nu }\,L_{n} &=  \hf^1_{n+1}+\hf^2_{n+1}  +
  {1\over  \nu }\, L^{\rm (f{}f)}_{n}\,=
  \hf^1_{n+1}+\hf^2_{n+1} + {C_{ij}\over 2}(\hh^i \hh^j)_n +
  \big({1\over \nu}-1\big)(\p \hh^3)_n\,,
\cr\cr
   \aw \,W_n  &=   \hf^{3}_{n+2}  +
  {1\over 2} \big(\kk-1\big) \,( \p\hf^{1} - \p \hf^{2})_{n+1}+
                     C_{2i} \, (\hh^i\,\hf^{1})_{n+1}
                    - C_{1i} \, (\hh^i\,\hf^{2})_{n+1}
                                  + \aw\, \Wf_n \,,
\cr\cr
  \aw \Wf_n &=
  C_{1i}C_{2j} (C_{1l}-C_{2l})\,(\hh^i\,(\hh^j\,\hh^l))_n
\cr\cr
            &\ + {1\over 2}\, \big(\kk-1\big) \,
              \left( C_{1i}\, (\hh^i\, \p\hh^1) -
              C_{2i}\, (\hh^i\,\p\hh^2) \right)_n
          \, + {1\over 6}\, \big(\kk-1\big)^2 \, (\p^2 \hh^1-\p^2 \hh^2)_n
                          \,.\cr}
\eqno(12)$$
 The standard  normalization [14] of the $W$ current is  recovered
choosing $\aw= - \nu^{-3/2} \sqrt{ 5c_\nu +22 \over 48}$ for the
 overall constant.

After appropriate identifications $\Tf$ and $\Wf$  reproduce
the free field realization of [14].
Both (12)
 and their (f{}f) parts  have the
 commutation relations of the Zamolodchikov
$\W_3$ algebra with conformal anomaly
$$
  c_\nu = 50 -24\,\Big( \nu + \kk \Big) \,.
\eqno(13)$$

The expressions for the reduced currents were computed in [4]
by directly solving the cohomological ``tic-tac-toe'' set of equations
and they coincide (up to some
numerical misprints) with the so defined (12).

The BRST operator implementing the constraints (7) is
$$
   Q^{\pb} =
   (e^3\,c^3)_{-1} + (e^2\,c^2)_{-1} - c^3_0 \,.
\eqno(14)$$
The hatted quantities now read
$$\eqalign{
  \he^1 &=e^1 + ( b^3 \,c^2 )\,,\qquad \he^2=e^2\,,\qquad \he^3=e^3\,,
\cr\noalign{\medskip}
  \hf^1 &=f^1 + (b^2 \, c^3 )\,,\qquad \hf^2=f^2\,,\qquad \hf^3=f^3\,,
\cr\noalign{\medskip}
  \hh^1 &= h^1- (b^2 \,c^2)+ (b^3 \,c^3) \,, \quad
  \hh^2  = h^2+2(b^2\,c^2)+(b^3\,c^3)\,.
\cr} \eqno(15)$$
The reduced quantum generators computed in [4] are again recovered
according to the rules of quantisation of the classical expressions (9),
 this time identifying the parameter $\zk$
with $\zk^{\pb}\equiv \kk-2= k+1\,.$ Using for simplicity
in the expansions of the fields the dimensions
inherited from the KM
algebra (i.e, $\triangle_{G^-}=2=2 \triangle_{G^+}\,,$
while the standard
half-integer  modes are recovered by a simple redefinition) one has

$$\eqalign{
    H_n  &= {1\over 3} (\hh^2_n - \hh^1_n) \,,
\quad     G^+_n  = \hf^1_n        \,,
\quad     G^-_n  = \hf^2_{n+1} + (\he^1\,\hh^2)_n + (\p\he^1)_n   \,,\cr\cr
   {1\over \nu}  L_n  &= \hf^3_{n+1} + (\he^1\,\hf^1)_n +
     {C_{ij}\over 2} (\hh^i\,\hh^j)_n
+({1\over \nu}-2) ({\p \hat h^3\over 2})_n  \,.
\cr}\eqno(16)$$
One readily checks that (16) generate the  $\W_3^{(2)}$ algebra
with conformal anomaly
$$   c_\nu = 25 - {6\over\nu} - 24 \, \nu\,. \eqno(17)$$

\medskip\noindent{\bf 5.\ }
We will consider modules  $\,\Omega_{\zl}$
that are tensor products of a $A_2^{(1)}\,$ Verma module with highest
weight vector $\V\,$ and ghost Fock module.
 To simplify notation we
will avoid indicating explicitly tensor products, thus assuming that
$\V\,$  is annihilated
by the positive modes of all $b^i(z)\,$
and the nonnegative modes of $c^i(z)\,$. Clearly any
singular vector in the module of  $A_2^{(1)}\,$ built on $V_{\zl}\,$
is a singular vector in  $\,\Omega_{\zl}\,$
and furthermore we can use equivalently  the hatted counterparts
of the three generating
elements of ${\tt N_-}$ to build these
vectors,   since $(b^i\,c^j)_0\,V_{\zl} = 0\,.$
The BRST charge $Q\,$ annihilates all singular vectors.
It is  immediate that this is also true for the positive modes of the
reduced generators (12) or (16), as well as for the zero mode $G^-_0$.
Let now  $V_{w \cdot \zl}\,$ be  some singular vector of weight $w \cdot \zl\,$
(including also the vacuum state).
For the zero modes in the (Z) case we have
$$
  L_0 \, V_{w \cdot \zl} =  \Lf_0 \, V_{w \cdot \zl}=
    h^{(2)}_{w \cdot \zl}   \, V_{w \cdot \zl} \,,\qquad
  W_0 \, V_{w \cdot \zl} = \Wf_0 \, V_{w \cdot \zl}
  = h^{(3)}_{w \cdot \zl} \, V_{w \cdot \zl} \,,
\eqno(18)$$
where
$$\eqalign{
  h^{(2)}_{\zl} &=
   {\nu \over 2}C_{ij}\, \zl (h_0^i)\,(\zl  -2\zk \zr)(h_0^j)
\,  =
\,   {\nu\over 2} \la\ol,\ol-2\zk\rho\ra \,,
\qquad \zk = \kk - 1 \,,
\cr
\aw   h^{(3)}_\zl &= C_{1i}\, C_{2j}\, (C_{1l}-C_{2l})\,
  \la\ol-\zk\rho,\za_i\ra \,
  \la\ol-\zk\rho,\za_j\ra \, \la\ol,\za_l\ra \,,\cr}
\eqno(19)$$
and they are invariant under the shifted action on $\zl$
 of the finite group generated by $w_0 w_1 w_0$ and  $w_0 w_2 w_0$.
This is equivalent to the
well known invariance under a $\zk$-shifted action of
the finite Weyl group on the projected weights,
i.e., if $\ol'-\zk\rho = w(\ol-\zk\rho)$ then
$ h^{(p)}_\zl =  h^{(p)}_{\zl'}$, $p=2,3$ and $w$ is
a word made of $w_1$ and $w_2$.

In the (PB) case   we have
$$
  L^{\pb}_0 \, V_{w \cdot \zl} = h^{\pb}_{w \cdot \zl}
   \, V_{w \cdot \zl} \,,\qquad
  H_0 \, V_{w \cdot \zl} = q_{w \cdot \zl} \, V_{w \cdot \zl} \,,
\eqno(20)$$
where
$$
  h^{\pb}_\zl = {\nu\over 2} \la\ol,\ol-\zk^{\pb}\rho\ra \,,  \qquad
  q_\zl =  (C_{2j}-C_{1j})\, \la\zl,\za_j\ra \,,
  \qquad\qquad \zk^{\pb} = \kk - 2 \,.
\eqno(21)$$
Now (21) are invariant under
the shifted action of $w_0$ on $\zl$ or equivalently -- under
 a ${1\over 2}\zk^{\pb}$-shifted action on the projected weights $\ol$ of
the reflection in the $\za_3$ direction, i.e., if
$\ol'-{1\over 2}\zk^{\pb}\rho
= w_{\za_3}(\ol-{1\over 2}\zk^{\pb}\rho)\,,$ then
$ h^{\pb}_\zl =  h^{\pb}_{\zl'}$ and $q_\zl=q_{\zl'}$.
In particular we can identify $V_{\zl^{}}$ with the highest weight state
 $| h^{(2)}_{\zl^{}}\,, h^{(3)}_{\zl^{}}\ra\,,$ or
  $|q_{\zl^{}}\,,h^{\pb}_{\zl^{}}\ra\,,$
 of a $\cW_3\,,$  or a $\cW^{(2)}_3$ Verma module.

\bigskip\noindent{\bf 6.}
Now we introduce  quantum analogues of the highest weight gauge fixing
transformations, which
 will be used as a tool to transform KM singular vectors into $\cW$
algebra ones.
Starting with the two simplest vectors in (2), corresponding to the
simple roots $\za_1,   \za_2\, $ of $A_2\,, $ it is clear that
  it is sufficient to have projections of
the ``full'' quantum gauge transformation along ``simple root directions''.
Thus
in the (Z) case the two (projected) transformations are a straighforward
generalization of the $A_1^{(1)}$ case [1], i.e.,
$$
  \Ri \equiv \Ri(\he^i_0) \qquad i=1,2\,.
\eqno(22)$$
where
$$
  \cR^{(i)}(u) = \odots \exp \Phi_i(-u) \odots
  %_{(-)}
     \,, \qquad
  %\p_u
   \Phi_i(-u) = C_{ij} \,\int du \,  \hh_{(-)}^j(-u) \,,
\eqno(23)$$
the subscript $\scriptstyle{(-)}$ denotes the holomorphic part
of the field and the $\odots\ \odots$ indicate that
in the expansion of the exponent the generators of the
gauge transformation $u(=\he^i_0)$
should come to the right of the modes of $\Phi$, i.e.,
more explicitly,
$$\eqalign{
  \cR^{(i)}(u) &= 1 + C_{ij}\hh^j_{-1}\,u +
  {1\over 2} \left( (C_{ij}\hh^j_{-1})^2 - C_{ij}\hh^j_{-2}\right)\,u^2
  + \dots  \cr\cr
  &= \sum_{k=0} \cR^{(i)}_{-k} \,u^k \,,  \quad {\rm where }
   \quad k\,\cR^{(i)}_{-k} = \sum_{l=0}^{k-1}\, (-1)^{k+l-1}\,
  \cR^{(i)}_{-l}\, C_{ij}\hh^j_{-k+l}\,, \, \,   \cR^{(i)}_{0} =1\,.\cr}
\eqno(24)$$

In the (PB) case the transformation in the $\za_1$ direction is the
identity (obvious since $G^+(z)=\hf^1(z)$) while in the $\za_2$
direction it is
$$
  \cR^{(2)} =  \odots \exp \he^1_{-1}\, e^2_0 \odots  \equiv \sum_{k=0}
{1\over k!} (\he^1_{-1})^k  (e^2_0)^k \,.
\eqno(25)$$

The quantum transformations have the following properties: \smallskip

{\it $\cR^{(i)}$ keeps all KM singular vectors invariant and
maps the states $V_t^{(i)} =
 (f^i_0)^{t}\,\V\,,$
 %$ i=1,2\,,$
  into the kernel of the BRST operator}.
{\it $\cR^{(i)}$ intertwines KM and $\cW$ algebra generators}.\smallskip

More precisely the last property
in the (Z) case takes the form (no summation in $i$)
$$\eqalign{
  \Ri \, \big( h^3_0 +2-\kk \big)\, f^{i}_0 \, V_t^{(i)}
&={1\over \nu}
  \sum_{p=1} \left(\left( \ze^i\,\aw \, \nu W_{-p}
        - {1\over 2}\big(1+\kk(2C_{ii}-1)\big)\,(\p L)_{-p}
 \right.\right.
\cr\cr
  & \hskip 1cm \left.\left.
  - {C_{ii}\over \nu}\, L_{-p}
 \right) \Ri
  + L_{-p}\ \Ri \  C_{ij}\, h^j_0 \right) \, (-e^{i}_0)^{p-1} \, V_t^{(i)} \,,
\cr}\eqno(26)$$
%where $V$ is any vector annihilated by all positive mode generators and
%by $c^a_0$, $a=1,2,3$, and an eigenvector of $h^3$.
where $\ze^1=1\,, \ze^2=-1$.
Its proof is rather
 lengthy, though straighforward, and
will be given in the detailed account of this work.
The idea of the proof is roughly
 the same as in [1] -- moving the gauge generator
$u(=\he^i_0)$ to the right produces the free field part (f{}f) of the reduced
generators while the remaining parts arise when we move to the right the
``gauge parameters'' (the modes of $\cR$ in the expansion in $u$).

In the (PB) case the intertwining property is
$$
 \cR^{(2)}\,  f^2_0  V_t^{(2)}  = \GM \,\cR^{(2)}\,  V_t^{(2)}
\eqno(27)$$
 proved by  straightforward computation.

Comparing  (6) and (8) with (24) and (25)
respectively, one sees that the latter
 can be viewed as some ``quantizations''
of the projections of the classical gauge transformations (taken in
an arbitrary representation).  As in [1] one can consider alternatively
the operators (23) with $u$ identified with an auxiliary
$sl(3)$ generator $t_+^i$ instead of $\he_0^i$.

\bigskip\noindent{\bf 7.}
Having the quantum gauge transformations now we can describe how
the KM singular vectors get transformed into $\cW$ algebra ones.
Again the arguments are generalization of the ones in [1].

Let us start with the case when the singular vector corresponds
to a single simple root $\za_1$ or $\za_2$, as in the first
equality in (2). In the (PB) case we have
$$
  (f^1_0)^{M_1}\,\V = (\GP)^{M_1}\,\V
   \quad{\rm if}\ M_1\in\dN
    \qquad{\rm or}\qquad
  (f^2_0)^{M_2}\,\V = (\GM)^{M_2}\,\V
   \quad{\rm if}\ M_2\in\dN\,,
\eqno(28)$$
the first being a trivial identity while the second is obtained by
applying repeatedly the relation (27)
and using  that $\cR^{(2)}$ leaves the
KM singular vectors invariant.
(Recall that our moding of $G^\pm$ (16) differs from the standard one.)
Thus we can identify
 the KM singular
vector $V_{w_{\za_i} \cdot \zl}\,, i=1\,,$ or $i=2\,, $
 with a singular vector
 (given explicitly by the r.h. sides of (28))
  in the $\cW^{(2)}_3$ Verma module.

In the (Z) case iterating the intertwining relation (26)
and using the properties of $\cR^{(i)}$ one gets
$$
   (- f^{i}_0)^{M_i}\, \V = \cO^{(i)}_{\zl} \, \V
   \qquad{\rm if} \quad M_i\equiv\la\zl+\rho,\za_i\ra\in\dN\,,
\eqno(29)$$
with
$$
   \cO^{(i)}_{\zl} = \cL^{(i)}_{M_i,0} +
  \sum_{k=1}^{M_i-1} \sum_{\{p_a\}_{a=1}^{k}}
  \cL^{(i)}_{M_i, p_{k}} \, \cL^{(i)}_{p_{k}, p_{k-1}} \,
  \dots \, \cL^{(i)}_{p_1, 0}\,,
\eqno(30)$$
where the second sum is over all $\{M_i > p_{k} > \dots > p_1 > 0\}$
and we have denoted
$$\eqalign{
  \cL^{(i)}_{t,t-p}\, &=\, {\prod_{a=1}^{p-1}
                           (t-a)(\la \zl+\rho,\za_i \ra + a - t )\over
                          \nu\, (t + \la \zl+\rho,\za_0 \ra )}\, \cdot
\cr\cr
  &\ \cdot\left(
  \ze^i \, \aw\,\nu \,W_{-p} +
  \Big( C_{ij} \la\zl,\za_j\ra - t + p - {2\over 3\nu}\Big)  L_{-p}
  - {1+3\nu\over 6\nu} \, (\p L)_{-p} \right) \,.
\cr}\eqno(31)$$
The $\cW_3$ singular vectors in the r.h.s. of (29) were obtained in
[7] using the method of ``fusion'' [15].
Iterating (26) for any $\, t=1,2,\dots, M_i\,,$ one actually
recovers also the basis elements of the matrix system equivalent
to
% the (scalar) vector
  $\, \cO^{(i)}_{\zl} \, \V\,.$
Note that in our approach the proof that   $\, \cO^{(i)}_{\zl} \,
\V\,$ is annihilated by the positive modes of the  $\cW_3$
currents is straightforward.

A remark is in order.
 In iterating (26) one can encounter examples when
the l.h.s. vanishes at some step.
 For an illustration consider the case $M^1=M^2=1$, then one has
to use (26) once, i.e., $t=0$. Now if $\nu=1$ one checks using
the explicit expression (12) for $W_{-1}$ that $W_{-1}\, \V =0
$ and furthermore both sides of (26) in this case vanish due to numerical
coefficients, thus not producing a nontrivial singular vector of $\cW_3$.
This phenomenon has to do with the existence of exceptional  $\cW_3$  modules
defined by indecomposable representations of $\{ L_0, W_0 \}$
(see e.g. [16]) and
we leave their analysis within our approach to a future investigation.

Next let us consider the example when both $M_1$ and $M_2$ are
positive integers. In this case also composite singular vectors
exist, i.e., we have that
$ (f^{1}_0)^{M_1+M_2}\, (f^{2}_0)^{M_2}\, \V\,\,$,
$ (f^{2}_0)^{M_1+M_2}\, (f^{1}_0)^{M_1}\, \V\,$ and
$ (f^{2}_0)^{M_1}\, (f^{1}_0)^{M_1+M_2}\, (f^{2}_0)^{M_2}\, \V$
together with the highest weight vector and the other two
singular vectors discussed above  form a hexagon of singular vectors.
The reduction of such composite vectors proceeds by reduction
of the separate factors. This is straightforward because (counting
from the right) the first, the first two, etc. factors also
produce singular vectors. For example in the (Z) case
$$
 (f^{2}_0)^{M_1}\, (f^{1}_0)^{M_1+M_2}\, (f^{2}_0)^{M_2}\, \V =
 \cO^{(2)}_{w_1w_2\cdot\zl} \, \cO^{(1)}_{w_2\cdot\zl} \,
  \cO^{(2)}_{\zl} \, \V \,,
\eqno(32)$$
 while in the
(PB) case one gets
$$
 (f^{2}_0)^{M_1}\, (f^{1}_0)^{M_1+M_2}\, (f^{2}_0)^{M_2}\, \V =
 (\GM)^{M_1} \,  (\GP)^{M_1+M_2} \,  (\GM)^{M_2} \V \,,
\eqno(33)$$
i.e., again  equalities of singular vectors.

The last series in (2), i.e., the vectors originating from the
affine (simple) root $\za_0$ do not survive under reduction.
Indeed starting with the (PB) case one can express
$e^3_{-1}=1+\{Q^{\pb},b_0^3\}$ and using that $Q^{\pb}$ annihilates the
vacuum state we have
$$     (e^3_{-1})^{{1\over \nu} -M^3}\,  \, \V = \V + Q^{\pb}\, \dots
  \V \,,
\eqno(34) $$
i.e., up to terms in the image of $Q^{\pb}$ we recover the vacuum
highest weight state which is in agreement with the invariance of (21).
(Note that since the modes $H_n, T_n, G_n^+\,, \, n > 0\,, \, G_m^-\,,
m\ge 0\,,$ annihilate the KM singular vectors $\V\,, V_{w_0 \cdot
\zl}\,,$ they also annihilate the $Q\,$ - exact term in the r.h.s. of (34).)
The result is even more trivial in the (Z) case,  since $e^3_{-1}=\{Q,b_0^3\}$,
and hence the whole singular vector is  $Q$ exact.

At the end we turn to the reduction of the general singular vectors
(see (1)). Postponing the detailed analysis
 here we only illustrate
the method on the example (3), where $M_1=m-{1\over \nu}\,, m\in \dN\,$.

The (PB) case is in complete analogy with the reduction of $A^{(1)}_1$
[1]. One has to combine the results above for the singular vectors
corresponding to the simple roots. Namely one transforms sequentially
the factors of  (3) starting from the left and
using (28) for the (nonaffine) directions $\za_1$ and $\za_2\,$.
 For the affine ($\za_0$) direction
 one
 uses (34) and recalling that
the reduced generators commute
with the BRST operator one gets
$$
  V_{w_1 w_0 w_2 w_0 w_1 \cdot \zl}
  = (\GP)^{m+\kk} \, (\GM)^{m} \, (\GP)^{m-\kk} \, \V  + Q^{\pb}\, \dots
  \V\,.
\eqno(35)$$
The first term in the r.h.s. provides a singular vector
in the $\cW^{(2)}_3$ Verma module.
One can make sense of the above
monomial with $G$'s raised
to complex powers as one does in the KM case
 (e.g., for the vector in (3)),
namely the structure is
such that the middle generator is to an integer power while
each successive pair (starting from the middle and going outwards) of
   surrounding generators
have powers adding to an integer -- commuting repeatedly from the
middle outwards
we get an ordinary polynomial.

The (Z) case is more subtle. To be able to carry rigourously the analysis
one has to make sense of (30) when $M_i$ is noninteger.
 For the analogous operator in the  Virasoro case  this was done in [17].
Let us assume that it is possible to
make an analytic continuation of (30)
to complex
powers of the nonaffine generators. Furthermore
for the reduction in the affine direction
we cannot use directly the argument above because raising
 the constraint
  $\, e^3_{-1} = \he^3_{-1}=\{Q, b_0^3\}$  to a complex
power is now ill defined.
On the other hand the invariance of (19) indicates that we should
consider the group of generators corresponding to $w_0 w_2 w_0$
in the middle of (3)
%or $w_0 w_1 w_0\,,$
 instead of considering the powers of $e^3_{-1}$ --
which correspond to $w_0$. Indeed
 casting the middle triple into an integer powers form
 and accounting for the properties of $Q$ produces
a sum of powers of the generator $e^1_{-1}\,$ and
of proper  bilinear ghost combinations
which add up to recover powers of its hatted counterpart
$\he^1_{-1} = 1 + \{Q, b_0^1\} \,$ (cf. (11)).
Thus we  get again
$1+\{Q,\dots\}$ (up to a numerical coefficient) and hence for the
full vector
$$
  V_{w_1 w_0 w_2 w_0 w_1 \cdot \zl} = N
  \cO^{(1)}_{w_0 w_2 w_0 w_1 \cdot \zl} \,   \cO^{(1)}_{\zl} \, \V
   + Q\dots\V \,,
\eqno(36)$$
where $N$ is a  constant.
%The first term in the r.h.s. is a singular vector at energy level
%$h^{(2)}_{\zl}+ 2m\,$ in the corresponding $\cW_3$ Verma module.

\bigskip
\bigskip\noindent{\bf Acknowledgements}
\medskip
We are indebted to V.K. Dobrev for useful discussions.
 The financial support and  the hospitality  of INFN,
Sezione di Trieste and SISSA, Trieste are    acknowledged.

This work was supported in  part by the Bulgarian Foundation for
Fundamental Research under contract $\Phi -11 - 91. $

%\vfill\eject
\bigskip\noindent{\bf References}
\medskip

\item{[1]}
    A.Ch. Ganchev and V.B. Petkova,
    Virasoro singular vectors via quantum DS reduction, preprint
    SISSA -74/93/EP , Trieste.

\item{[2]} %[Z]
A.B. Zamolodchikov, {\it Theor. Math. Phys.} {\bf 65} (1988) 1205.

\item{[3]} %[P] [B]
A. Polyakov, {\it Int. J. Mod. Phys.} {\bf A5} (1990) 833.

M. Bershadsky, {\it Comm. Math. Phys.} {\bf 139} (1991) 71.

\item{[4]} %[BT2]
  J. Boer and T. Tjin, The relation between quantum $\W$ algebras
  and Lie algebras, preprint THU - 93/05, ITFA - 02 - 93.

\item{[5]}% [DS]
V. Drinfeld and V. Sokolov, {\it J. Sov. Math.} {\bf 30} (1984) 1975.

\item{[6]} %[BFFOW]
J. Balog, L. Feh\'er, P.Forgac, L. O$'$ Raifeartaigh and A. Wipf,
 {\it Ann. Phys.} {\bf B203} (1990) 76;

 F.A. Bais, T. Tjin, P. van Driel,
    {\it Nucl. Phys.} {\bf B357} (1991) 632.

\item{[7]} %[BW]
P. Bowcock and G.M.T. Watts, {\it Phys. Lett.} {\bf B297} (1992) 282.

\item{[8]} %[FORTW]
  L. Feh\'er,  L. O$'$ Raifeartaigh,
O. Ruelle, I.Tsutsui and A. Wipf,
 {\it Phys. Rep.} {\bf 222} (1992) 1.

\item{[9]} %[BouwkS]
   P. Bouwknegt and K. Schoutens, {\it  Phys. Rep.} {\bf 223} (1993) 183.

\item{[10]} %[KK]
 V.G. Kac and D.A. Kazhdan, {\it Adv. Math.} {\bf 34} (1979) 97.
\item{[11]}% [MFF]
   F.G. Malikov, B.L. Feigin and D.B. Fuks, {\it Funct. Anal.
   Prilozhen.} {\bf 20}, no.  {\bf 2} (1987) 25.

\item{[12]} %[T]
K. Thielemans,
{\it Int. J. of Mod. Phys.} {\bf C2} (1991) 787.

\item{[13]} %[FF]
   B. Feigin and E. Frenkel, {\it Phys. Lett.} {\bf B246} (1990) 75.

 \item{[14]} %[FZ]
 V.A. Fateev, A.B. Zamolodchikov,
 {\it Nucl. Phys.} {\bf B280} (1987) 644.

\item{[15]} %[BFIZ]
     M. Bauer, Ph. Di Francesco, C. Itzykson and J.-B. Zuber,
    {\it Nucl. Phys.} {\bf B362} (1991) 515.

\item{[16]} %[W]
  G.M.T. Watts,
 {\it Nucl. Phys.} {\bf B326} (1989) 648;
\hfill
\break
 %[BMP]
P. Bouwknegt, J. McCarthy and K. Pilch, On the BRST structure of $\W_3$
gravity coupled to $c=2$ matter, preprint USC-93/14, ADP-93-203/M17.

\item{[17]}% [K]
    A. Kent, {\it Phys. Lett.} {\bf B273} (1991) 56.

\bye